\documentclass[preprint,showpacs,preprintnumbers,amsmath,amssymb]{revtex4}

\usepackage{graphicx}
\usepackage{dcolumn}
\usepackage{bm}

\newcommand{\apjl}{Astrophys. J.}
\newcommand{\mnras}{Mon. Not. R. Astron. Soc.}
\begin{document}

\preprint{OU-TAP}

\title{Magnetic field generation by the Weibel instability at
temperature gradients in collisionless plasmas}

\author{Yutaka Fujita}
 \email{fujita@vega.ess.sci.osaka-u.ac.jp}
\affiliation{%
Department of Earth and Space Science, Graduate School of Science, Osaka
 University, Toyonaka, Osaka 560-0043, Japan
}

\author{Tsunehiko N. Kato}%
\affiliation{%
National Astronomical Observatory, Osawa 2-21-1, Mitaka, Tokyo 181-8588, 
Japan
}%

\author{Nobuhiro Okabe}
\affiliation{
Astronomical Institute, Tohoku University, Sendai 980-8578, Japan
}%

\date{\today}

\begin{abstract}
 The Weibel instability could be responsible for the generation of
 magnetic fields in various objects such as gamma-ray bursts, jets from
 active galactic nuclei, and clusters of galaxies. Using numerical
 simulations, the development of the Weibel instability at a temperature
 gradient is studied. It is found that current sheets are first
 generated at the gradient, and then they are rounded off and turn into
 current filaments. During this process, return currents are generated
 around the filaments and they prevent filaments from merger. The
 magnetic fields around the filaments persist at least until $t\sim
 8000/\omega_p$, where $\omega_p$ is the plasma frequency, and it is
 very likely that they survive for a much longer time.
\end{abstract}

\pacs{Valid PACS appear here}
\maketitle

\section{Introduction}

The question of the origin of magnetic fields in the Universe is one of
the most challenging problems in modern astrophysics. One of the
fascinating ideas is that the fields are generated by the Weibel
instability, a plasma instability in a collisionless plasma
\cite{wei59}. This instability is driven by the anisotropy of the
particle velocity distribution function (PDF) of plasma. 
When the PDF is anisotropic, currents and then magnetic fields are
generated in the plasma so that the plasma particles are deflected and
the PDF becomes isotropic \cite{med99}. Through this process, the free
energy attributed to the PDF anisotropy is transferred to magnetic field
energy. This instability does not need seed magnetic fields. It can be
saturated only by nonlinear effects, and thus the magnetic fields can be
amplified to very high values.  In particular, the generation of
magnetic fields at shocks through the instability has been studied by
many authors, because the PDF is anisotropic at shocks
\cite{med99,kaz98,nis03,sil03,sai04,fre04,fuj05,med06}. 

However, the magnetic fields generated by the Weibel instability at a
shock may not survive for a long time. At a shock, current filaments are
created and magnetic fields are generated around them
\cite{sil03,fre04}. As the filaments merge together by the magnetic
force between them, the magnetic field strength increases
\cite{fre04,med05}. However, the magnetic field should be saturated when
the current strength reaches the Alfv\'en current \cite{kat05}, which is
the maximum current allowed by the self-generated magnetic field
\cite{alf39}. The Alfv\'en current is given by
\begin{equation}
 I_{\rm A} = (m c^3/q)\gamma\beta\:,
\end{equation}
where $m$ and $q$ are the mass and charge of a particle, respectively,
$\beta$ is the mean velocity of particles normalized by the speed of
light $c$, and $\gamma=1/\sqrt{1-\beta^2}$. After the saturation,
magnetic field strength, $B$, would decrease as the filament size $r$
increases through mergers, because $B\propto I_A/r$ and $I_A$ is
constant, although current numerical simulations cannot fully deal with
this phase because of the limitation of the simulation box size.  The
timescale of the filament mergers before the saturation is $\sim
10\:\omega_p^{-1}$, where $\omega_p$ is the plasma frequency
\cite{med05}. If the timescale after the saturation is the same as that
before the saturation, the magnetic fields would rapidly fade away.

Another site where the Weibel instability could be effective is
temperature gradients, where the PDF is also anisotropic \cite{oka03}.
One example of such temperature gradients is cold fronts observed in
clusters of galaxies \cite{oka03,hat05}.
In this paper, we present the results of numerical simulations of
electron-positron plasma performed to investigate the long-term
evolution of the Weibel instability at a temperature gradient. 
We emphasize that in our simulations,
Coulomb collisions are ineffective (collisionless plasma), which is
different from the assumption of Ref.~\cite{oka03}.

\section{Models}

We performed numerical simulations of an electron-positron plasma at a
temperature gradient.  The simulation code used is a relativistic,
electromagnetic, particle-in-cell code with two spatial and three
velocity dimensions, which was developed based on a general description
by Ref.~\cite{bir91}.
The code is a momentum conserving code.  Using the code, we solve the
Maxwell equations (in Gaussian units):
\begin{equation}
 \frac{1}{c}\frac{\partial \bm{E}}{\partial t}
=\nabla\times\bm{B}-\frac{4\pi}{c}\bm{J}\:,
\end{equation}
\begin{equation}
 \frac{1}{c}\frac{\partial \bm{B}}{\partial t}
=-\nabla\times\bm{E}\:,
\end{equation}
\begin{equation}
 \nabla\cdot\bm{E}=4\pi\rho\:, \hspace{10mm}\nabla\cdot\bm{B}=0\:,
\end{equation}
where $\bm{E}$ is the magnetic field, $\bm{B}$ is the magnetic field,
$\bm{J}$ is the current density, and $\rho$ is the charge density.
We also solve the equation of motion for each particle:
\begin{equation}
 \frac{d\bm{p}}{dt}=q\left(\bm{E}+\frac{\bm{p}\times\bm{B}}
{\gamma m_e c}\right)\:,
\end{equation}
where $\bm{p}$ is the momentum of a particle, and $m_e$ is the electron
mass. The simulations are performed on a $1024\times 512$ grid (the axes
are labeled as $x$ and $y$, respectively) with a total of 20 million
particles.  Temporal and spatial scales in the simulations are
normalized to the inverse electron plasma frequency $\omega_p^{-1}=(4\pi
n_{e0} e^2/m_e)^{-1/2}$ and the collisionless skin depth
$\lambda_e=c/\omega_p$, where $n_{e0}$ is the average initial electron
or positron density, $-e$ is the electron charge, 
and $c$ is the speed of light, which is the unit of velocity in our
simulation code. The units of mass and charge are the electron mass
$m_e$ and the absolute value of the electron charge $e$,
respectively. In these normalized units, the box size is $160\times
80$. For electromagnetic fields, we adopt a periodic boundary
condition. For the $x$ direction, we set walls at $x=20$ and 140. The
regions, $0<x<20$ and $140<x<160$ are used for Joule dissipation of
electromagnetic waves so that electromagnetic waves do not enter from
the other side.

For particles, we adopt a periodic boundary condition for the $y$
direction. On the other hand, for the $x$ direction, we adopt reflection
boundary conditions with 'heat walls'.  Particles that hit the wall at
$x=20$ have a thermal velocity of $\sigma_L=0.1$. That is, the reflected
particles have the Maxwellian velocity distribution with the deviation
of $0.1$. At the wall of $x\approx 140$, reflected particles have the
one with the deviation of $\sigma_R=0.5$. For the right wall, we
randomly choose the reflection point in $137<x<140$ to avoid the
formation of artificial structures in the plasma at the wall.
Because of these boundary conditions, the total energy of the system
does not conserve.

Initially, the plasma in the simulation box has no magnetic fields and
is `isobaric', that is, the square of the thermal velocity
(`temperature') times the density is constant. At the left ($x=20$) and
right ($x=140$) boundaries, the plasma has thermal velocities of
$\sigma_L=0.1$ and $\sigma_R=0.5$, respectively.  For $20<x<140$, the
temperature changes linearly as $x$ increases.  Thus, the density on the
left side is higher.

It is to be noted that in another simulation we have calculated the
plasma evolution when two plasmas with different temperatures but with
the same pressure are bordered at $x=80$. The results are qualitatively
the same as those shown below.

\section{Results}

Fig.~\ref{fig:history} shows the evolution of the magnetic energy,
$W_B$, in units of the total energy in the box at $t=0$ (the sum of the
kinetic energies of all particles). 
For comparison, the total particle energy, $E_P$, is shown.  The total
particle energy decreases because of the cooling at the left wall as
well as the generation of magnetic fields.

\begin{figure}
\includegraphics[width=84mm]{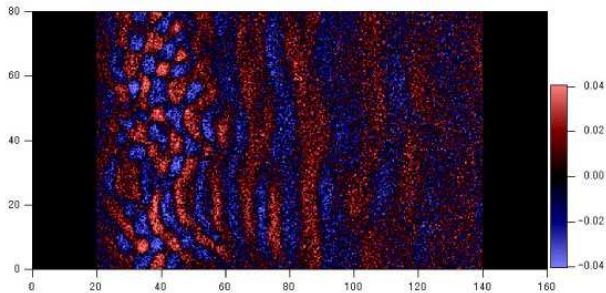} \caption{(Color online). The current
density in the $z$-direction ($J_z$) at $t=400$ in units of
$n_{e0}\: e c$.}  \label{fig:Jz400}
\end{figure}

The magnetic energy rapidly increases at $t\lesssim 400$. Because of the
initial density gradient, particles as a whole move in the right
direction. In particlar, those with large velocities move fast.  Thus,
on the frame moving with these particles, the effective temperature of
the particles is smaller in the $x$-direction than those of the other
directions, which develops the Weibel instability. This initial stage of
the instability will be studied in detail
elsewhere. Fig.~\ref{fig:Jz400} shows the current density in the
$z$-direction ($J_z$) at $t=400$. Current sheets are seen at $x\lesssim
80$. Strong magnetic fields are formed around these current sheets. For
$x\gtrsim 80$, sheets are still developing. On the other hand, the
current density in the $x$-direction ($J_x$) is much weaker than $J_z$
and does not contribute to the formation of magnetic fields.

\begin{figure}
\includegraphics[width=74mm]{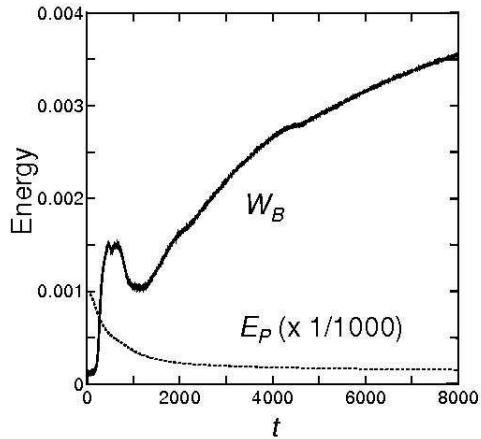} \caption{The evolution of the
magnetic energy ($W_B$: solid) and the total particle energy ($E_P$:
dotted) in units of the total energy in the box at $t=0$.}
\label{fig:history}
\end{figure}

Note that contrary to the Weibel instability at the temperature
gradient, current sheets are not created at shocks. While the thermal
velocity of particles is larger in the `directions' perpendicular to the
temperature gradient in the calculation presented here, it is larger in
the `direction' of the shock normal at a shock front \cite{med99}. This
is the reason why only current sheets (two-dimensional structures) are
generated in the former. In the latter, only current filaments are
formed in the direction of the shock normal.

At $t\sim 400$, $W_B$ reaches its local maximum. We found that the
thickness of individual current sheets at $x\lesssim 60$ in
Fig.~\ref{fig:Jz400} is comparable to the gyroradius of particles. Note
that for current filaments, if the radius of the individual filaments is
comparable to the gyroradius of particles, the current strength is to be
the Alfv\'en current \cite{kat05}. It has been shown that the magnetic
field strength should reach its maximum at that time \cite{kat05}.

\begin{figure}
\includegraphics[width=84mm]{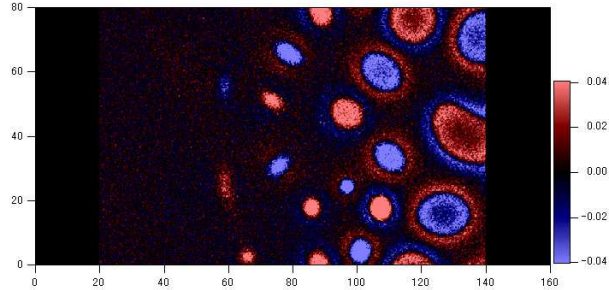} 
\caption{The same as Fig.~\ref{fig:Jz400} but for
 $t=5000$.}
\label{fig:Jz5000} 
\end{figure}

At $650\lesssim t\lesssim 1100$, the mixing of hot and cold plasmas
proceeds, and $W_B$ decreases. However, for $t\gtrsim 1100$, $W_B$
starts increasing again. In this period, the current sheets are rounding
and turning into `filaments', because the sheets are unstable. 
Since the sheets are perpendicular to the temperature gradient, the
filaments should also be perpendicular to that direction. In this
process, the cross sections of the sheets decrease.  The magnetic field
strength around the filaments slowly increases through the rounding and
shrinking. The increase of the magnetic fields produces electromotive
forces around the filaments, which induce return currents around the
filaments. The return currents are clearly seen around filaments at
$t\gtrsim 1500$ (Fig.~\ref{fig:Jz5000}). 

Since the return currents shield the magnetic fields outside the
filaments, the magnetic force between the filaments is significantly
reduced. At this stage, the filaments stop merging.  While the filaments
at large $x$ do not evolve, the sheets continue to round at smaller $x$,
and $W_B$ gradually increases even at $t\gtrsim 2000$
(Fig.~\ref{fig:history}). Fig.~\ref{fig:Jz5000} also shows that current
density is not uniform in a filament. The current density is small at
the center of the filament and thus each filament appears to be `a
current tube'.

In Fig.~\ref{fig:Jz5000}, filaments are seen only at $x\gtrsim 80$. The
thermal velocity outside the filaments does not much depend on $x$ and
is $\sim \sigma_L=0.1$. Particles with large velocities come from the
right wall. 
However, the magnetic field around the filaments prevents those
particles from moving to the left wall. Thus, the PDF anisotropy is
larger at larger $x$, which is the reason that filaments develop at
larger $x$.

In our simulations, the particles are merely reflected and redistributed
thermally at the right and left walls; they are not uniformally fed into
the system. In spite of this, the walls do not create artifacts. This
may be because particles in low and high density regions bordered on the
same wall are well-separated by magnetic fields and they are not mixed
together. It would be interesting to compare the results here with those
of the simulations in which particles are uniformally fed into the
system.

\section{Discussion}

\begin{figure}
\includegraphics[width=84mm]{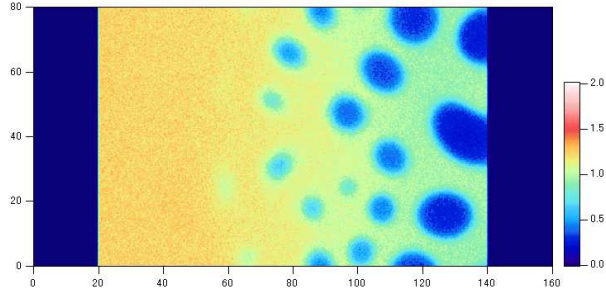} \caption{(Color online). The number
 density of electrons at $t=5000$ in units of $n_{e0}$.}  \label{fig:ne}
\end{figure}

\begin{figure}
\includegraphics[width=84mm]{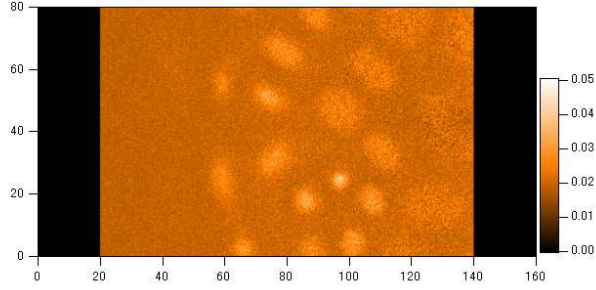} \caption{(Color online). The
kinetic energy density of particles (electrons and positrons) at
$t=5000$ in units of $n_{e0}m_e c^2$.}  \label{fig:U_prt}
\end{figure}

\begin{figure}
\includegraphics[width=84mm]{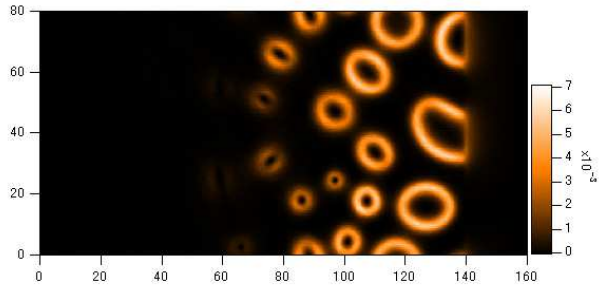} \caption{(Color online). The
magnetic energy density, $U_B$, at $t=5000$ in units of $n_{e0}m_e
c^2$.}  \label{fig:U_B}
\end{figure}

We studied the structure of the stable filaments seen at $x\gtrsim 100$
in Fig.~\ref{fig:Jz5000}. In the filaments close to the right wall, the
average drift velocity of particles in the $z$-direction is $|\beta_{\rm
in,z}|\sim 0.05$, while their thermal velocity is $\sigma_{\rm in}\sim
0.16$. This means that the filaments are `hot beams'. Figs.~\ref{fig:ne}
and~\ref{fig:U_prt} show the number density and the kinetic energy
density of particles at $t=5000$, respectively. For the filaments at
$x\gtrsim 100$, while the number density is small inside them, the
kinetic energy density is almost the same between the inside and the
outside of the filaments.  Fig.~\ref{fig:U_B} shows the magnetic energy
density ($U_B$), and demonstrates that magnetic fields are generated
only at the surfaces of the filaments.
There is a relation of $U_B=(1/2)(B/B_\star)^2$, where $B_\star =
c\sqrt{4\pi n_{e0} m_e}$. The magnetic field strength is determined so
that the gyroradius of particles with the velocity of $\sigma_{\rm
in}\sim 0.16$ is comparable to the thickness of the current tubes. This
magnetic field confines hot particles within the tubes.

Based on these facts, we can analyze the structure of the magnetic
fields as follows. 
We use the Gaussian units here. We define `thermal pressure' as
\begin{equation}
 P=2 n_e m_e \sigma^2 c^2\:,
\end{equation}
where $n_e$ is the density, and $\sigma$ is the thermal velocity of
particles. The factor of two comes from the fact that there are
electrons and positrons. Our simulations showed that the thermal
velocities inside and outside a filament are $\sigma_{\rm in}\lesssim
\sigma_R$ and $\sigma_{\rm out}\gtrsim \sigma_L$ ($\sigma_{\rm
in}\gtrsim \sigma_{\rm out}$), respectively. Fig.~\ref{fig:U_prt}
suggests that
\begin{equation}
 P_{\rm in}\approx P_{\rm out}\:,
\label{eq:balance}
\end{equation}
where $P_{\rm in}$ and $P_{\rm out}$ are the thermal pressures inside
and outside a filament, respectively. Equation~(\ref{eq:balance}) means
that
\begin{equation}
 n_{\rm in}\sigma_{\rm in}^2\approx n_{\rm out}\sigma_{\rm out}^2\:,
\label{eq:balance2}
\end{equation}
where $n_{\rm in}$ and $n_{\rm out}$ are the electron densities inside
and outside a filament, respectively. 

Since the gyroradius of particles inside a filament is comparable to
the thickness of the current tube ($\Delta$), the magnetic field
strength is given by
\begin{equation}
 B\approx \eta
\frac{m_e c^2 \sigma_{\rm in}}{\Delta e}\:,
\label{eq:mag}
\end{equation}
where $\eta$ is the ratio of the thickness to the gyroradius. On the
other hand, the magnetic field strength is also given by $B=2I/(rc)$,
where $I$ is the current in the filament, and $r$ is the radius of the
filament (or the tube). If we assume that the average drift velocity of
particles in a filament is given by $|\beta_{\rm in,z}|=\xi \sigma_{\rm
in}$, the current is $I=4\pi n_{\rm in} e \beta_{\rm in,z} c r
\Delta$. Thus, the magnetic field strength is
\begin{equation}
 B=8\pi \xi n_{\rm in} e\sigma_{\rm in} \Delta\:.
\label{eq:mag2}
\end{equation}
Combining equations~(\ref{eq:balance2}), (\ref{eq:mag}), and
(\ref{eq:mag2}), we obtain
\begin{equation}
 B\approx \sqrt{2\eta\xi}B_\star\sigma_{\rm out} \:,
\label{eq:mag3}
\end{equation}
and
\begin{equation}
 \Delta\approx \sqrt{\frac{\eta}{2\xi}}\lambda_e 
\frac{\sigma_{\rm in}}{\sigma_{\rm out}}\:.
\label{eq:gyro}
\end{equation}
We used the relation $n_{\rm out}\approx n_{e0}$. In our simulations, we
observed $\sigma_{\rm in}=0.16$, $\sigma_{\rm out}=0.08$, and $\xi =0.3$
for the filaments close to the right wall. For these values and
$\eta=2$, equations~(\ref{eq:mag3}) and~(\ref{eq:gyro}) predict that
$B/B_\star \approx 0.09$ and $\Delta/\lambda_e\approx 4$, which are
consistent with the results of the simulations (Fig.~\ref{fig:U_B}). The
above equations alone cannot determine the radius of a filament $r$. It
may depend on the initial conditions of plasma.

\begin{figure}
\includegraphics[width=84mm]{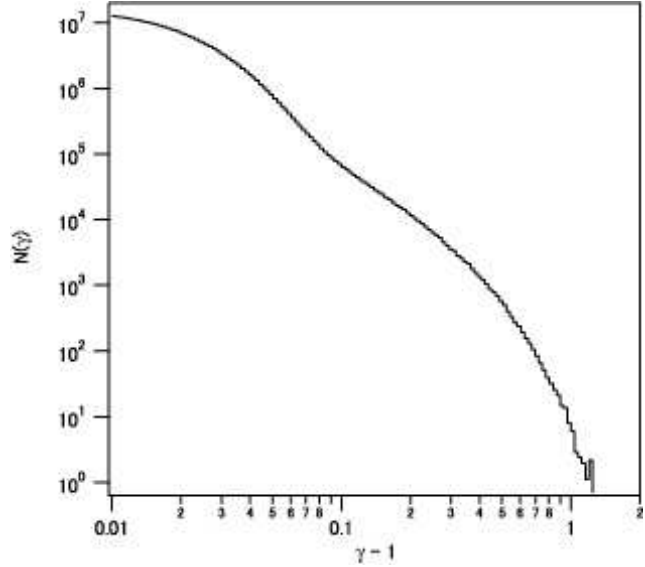} \caption{The histogram of
 $\gamma-1$ (the kinetic energy of particles in the unit of $m_e c^2$)
 at $t=8000$.}  \label{fig:gam}
\end{figure}

Although we finished the simulations at $t\sim 8000$, we expect that
$W_B$ continues to increase for $t\gtrsim 8000$, because of the rounding
and shrinking of sheets at smaller $x$. On the other hand, the filaments
at larger $x$ will remain the same. 
Fig.~\ref{fig:gam} shows the histogram of $\gamma-1$ (the kinetic energy
of particles in the unit of $m_e c^2$) at $t=8000$. It shows that
particles consist of those with the thermal velocity of $\sim\sigma_L$
and those of $\sim\sigma_R$. A sign of particle acceleration cannot be
found.

Our simulations are two-dimensional in space. Thus, we cannot deal with
three-dimensional deformations of current filaments, which might be
caused by the Kink instability or filament mergers. However, for the
mergers, if the time-scale of the formation of return currents is
smaller than that of the mergers, the results presented here would not
much change. For filaments that have not been shielded by return
currents, it would take a longer time to merge together, because the
filaments would not be aligned and the magnetic interaction among them
would be weaker in a three-dimensional space. Thus, filament mergers
would be further prohibited.  On the other hand, while our simulations
are two-dimensional in space, they are three-dimensional in velocity
space. Therefore, the resultant velocity distribution of particles would
not be much different from that in full three-dimensional simulations.

We have shown that magnetic fields generated at a temperature gradient
could survive as long as the gradient exists. The essence is the
generation of current sheets, which later round and turn into
filaments. Return currents created in this process shield the filaments
and prevent them from mergers.  It would be interesting to study this
instability for electron-proton plasma. Assuming that protons and
electrons have the same temperature, thermal velocity of the protons is
smaller than that of the electrons by a factor of $\sqrt{m_e/m_p}$,
where $m_p$ is the proton mass. Since $B_\star$ for protons is larger
than that for electrons by a factor of $\sqrt{m_p/m_e}$, the resultant
magnetic fields would not be much different from that for
electron-positron plasma (see Eq.~[\ref{eq:mag3}]), although it must be
confirmed by numerical simulations.  It would also be interesting to
apply the results to actual objects in the Universe, such as gamma-ray
bursts, supernova remnants, and clusters of galaxies. However, the
temperature gradients in them are much smaller than that studied in this
paper, which might make some difference. Moreover, the coherent scale of
the magnetic fields generated by the Weibel instability is much smaller
than astrophysical scales. Thus, some mechanisms that increase the scale
would be required \cite{fuj05}.  In the future, we would like to
challenge these problems.

\begin{acknowledgments}
Y.~F.\ is supported in part by a Grant-in-Aid from the Ministry of
Education, Culture, Sports, Science, and Technology of Japan
(17740182). 
\end{acknowledgments}

\appendix

\newpage


\end{document}